\documentclass[twocolumn,showpacs,preprintnumbers,amsmath,amssymb]{revtex4}

\usepackage{graphicx}
\usepackage{dcolumn}
\usepackage{bm}

\begin{document}


\title{Quantum phases of bosons in double-well optical lattices}
\author{I. Danshita$^{1,2}$}
\author{J. E. Williams$^3$}
\author{C. A. R. S\'a de Melo$^{1,4}$}
\author{C. W. Clark$^1$}
\affiliation{
{$^1$Joint Quantum Institute, National Institute of Standards and Technology, and University of Maryland, Gaithersburg, Maryland 20899, USA}
\\
{$^2$Department of Physics, Waseda University, 3-4-1 \=Okubo, Shinjuku, Tokyo 169-8555, Japan}
\\
{$^3$Wolfram Research, Inc., Champaign, IL 61820, USA }
\\
{$^4$School of Physics, Georgia Institute of Technology, Atlanta, Georgia 30332, USA}
}

\date{\today}

\begin{abstract}
We study the superfluid to Mott insulator transition of bosons in a two-legged ladder optical lattice, of a type accessible in current experiments on double-well optical lattices.
The zero-temperature phase diagram is mapped out, with a focus on its dependence upon interchain hopping and the tilt between double wells.
We find that the unit-filling Mott phase exhibits a non-monotonic 
behavior as a function of the tilt parameter, producing a reentrant phase transition
between Mott insulator and superfluid phases.
\end{abstract}

\maketitle
Optical lattices loaded with ultra-cold atoms provide the opportunity to study quantum phases of many-particle systems
because of their unprecedented degree of controllability~\cite{rf:zoller}.
Presently the lattice depth, dimensionality, geometry, and filling factor can all be reasonably controlled. 
While one of the first examples of this degree of control was the experimental observation of the superfluid (SF)-to-Mott insulator (MI) transition in three-dimensional cubic optical lattices
as a function of the lattice depth~\cite{rf:greiner},
tetragonal and orthorhombic optical lattices can also be produced by deepening the optical potential along desired directions~\cite{rf:ian,rf:stoeferle}.

\begin{figure}[tb]
\includegraphics[width=7cm, height=3.2 cm]{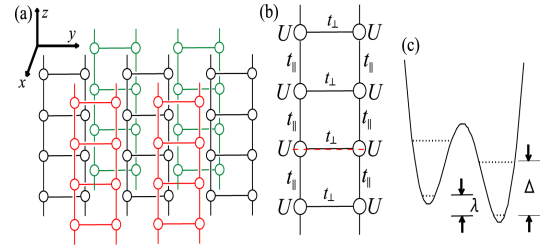}
\caption{\label{fig:twoleg}
(a) Schematic picture of a 3D configuration of a double-well lattice. Circles represent sites, 
and solid lines represent connections via the hopping between sites.
(b) Schematic picture of the two-leg ladder.
(c) Double-well potential corresponding to the cross section for the dashed line in (a). Dotted lines represent the energy levels for each well.
}
\end{figure}

More recently, possibilities for control have expanded with the experimental realization of double-well optical lattices.
Control of the polarization of the laser beams allows for the production of lattices with a base in two and three dimensions~\cite{rf:jenni}
as illustrated in Fig.~\ref{fig:twoleg}a, where Bose atoms ($^{87}{\rm Rb}$) have been successfully trapped.
In particular, one can create a one-dimensional double-well optical lattice corresponding to a two-leg ladder structure shown in Fig.~\ref{fig:twoleg}b by increasing the long period of the double-well optical lattice.
In standard condensed-matter physics, a few compounds, such as 
vanadyl pyrophosphate $({\rm VO})_2 {\rm P}_2 {\rm O}_7$~\cite{rf:johnston} and some cuprates like ${\rm Sr}{\rm Cu}_2{\rm O}_3$~\cite{rf:azuma},
have such two-leg ladders in their crystalline structure, and
they display much of the interesting physics encountered in general
ladder systems, associated with the interplay between spin-gapped and superconducting states~\cite{rf:dagotto}.
However, conventional condensed-matter systems come with fixed dynamical and structural parameters, 
while the flexible variability of optical lattices offers the prospect of exploring the full parameter space.
Moreover, the particles confined in the current double-well optical lattices are bosonic atoms, 
in contrast to conventional condensed-matter systems, where electrons (fermions) dictate the quantum phases.

In this paper, we study the zero-temperature phase diagram of bosons in double-well optical lattices.
We focus on the case of a two-legged ladder, where analytical and numerical progress can be made; in particular, we apply the time-evolving block decimation (TEBD)~\cite{rf:vidal1,rf:vidal2} method to such ladder systems.
We show that the phase diagram changes dramatically as a function of the chemical potential $\mu$,
the intrachain (interchain) hopping $t_{\parallel}$ ($t_{\perp}$), 
the on-site repulsion $U$ and the tilt $\lambda$ of the double-well, which are indicated in Figs.~\ref{fig:twoleg}b and \ref{fig:twoleg}c.
For $\lambda = 0$ and different ratios $t_{\perp}/U$, 
Mott phases with half-integer (in addition to integer) filling factors 
emerge in the phase diagram of $\mu/U$ versus $t_{\parallel}/U$ for small 
ratios of $t_{\parallel}/U$. 
As $t_{\perp}/U$ increases, the half-filling MI phase becomes larger and the unit-filling one becomes smaller.
For fixed ratio $t_{\perp}/U$, and different values of $\lambda/U$, we also obtain the $\mu/U$ versus $t_{\parallel}/U$ 
phase diagram which reveals a reentrant phase transition for the unit-filling MI
induced by the tilt $\lambda$.
The reentrant phase transition can be attributed to the development of coherence in each double well 
in the vicinity of $\lambda = U$ which drives the system into the SF phase.
Finally, we also calculate the critical points for the MI-to-SF transition at half and unit fillings.

To describe the physics discussed above, we introduce the Bose-Hubbard model for the double-well ladder
\begin{equation}
\label{eq:hamil}
  H = \sum_{i}[ H^{\rm dw}_i - t_{\parallel}\sum_{\eta\in\{L,R\}}(a^{\dagger}_{i+1,\eta} a_{i,\eta}\!+\!{\rm h.c.})],
\end{equation}
where $H^{\rm dw}_i$ represents the double-well Hamitonian for a given ladder index $i$ and is given by
\begin{eqnarray}
\label{eq:double-well-hamil}
H^{\rm dw}_i &=& \sum_{\eta}[\frac{U}{2} \hat{n}_{i,\eta} (\hat{n}_{i,\eta}-1)-\mu\hat{n}_{i,\eta}]\nonumber\\
  &&-t_{\perp}\!( a^{\dagger}_{i,R} a_{i,L}\!+\!{\rm h.c.})
  + \frac{\lambda}{2}(\hat{n}_{i,L}-\hat{n}_{i,R}),
\end{eqnarray}
$a^{\dagger}_{i,\eta}$ creates a boson at the lowest level localized on the left (right) of 
the $i$-th double-well when $\eta=L\;(R)$.
We do not include the effect of the harmonic trapping potential.
We assume that all the parameters are sufficiently small compared to the energy gap $\Delta$ between the first and second levels of each site.
Furthermore, all parameters of $H$ are controllable in experiments~\cite{rf:jenni},
and thus we begin our discussion by analyzing the limit of $t_{\parallel} = 0$.

{\it Integer and Half-Integer Mott Phases:} When $t_{\parallel}=0$ 
several MI phases emerge.
In this case, the Hamiltonian Eq. \!(\ref{eq:hamil}) reduces to $H = \sum_i H_{i}^{\rm dw}$.
One can obtain the eigenenergy $E^{(0)}(n,j)$ and eigenstate $\vert \Phi_{n,j}\rangle = \sum_{n_{L}=0}^{n}C_{n_{L}}(n,j)|n_{L},n-n_{L}\rangle$ of $H_i^{\rm dw}$,
where $n$ is the number of atoms in each double well
and the quantum number $j$ is a non-negative integer less than $n+1$ ($j=0,1,...,n$).
$|n_{L},n_{R}\rangle$ is the Fock state with $n_{L}$ ($n_{R}$) atoms on the left (right) of each double well.
When $n$ is even (odd), the filling factor $\nu$ of the MI phases is integer (half-integer).
Although there exists a MI phase for each value of $n$, we focus here only on the $\nu = 1, 1/2$ phases.

We consider first the case of symmetric double wells ($\lambda=0$) and discuss two limiting situations corresponding 
to $t_{\perp}\gg U$ and $t_{\perp}\ll U$. 
For $t_{\perp}\gg U$, the anti-bonding single particle state of the double-well is pushed to energies much higher than $U$, and only the
bonding single particle state and the lowest energy two-particle state are important. 
Therefore, in this limit, the $\nu = 1/2$ $(\nu = 1)$ phase can be mapped into the 
unit (double)-filling MI phase for a 1D lattice with an effective hopping $t_{\parallel}$ 
and an on-site repulsive interaction $U/2$.
For $t_{\perp}\ll U$, the $\nu = 1$ MI phase approaches that of a 1D lattice (two independent filling-one chains), and the width of the $\nu=1$ MI phase on the $\mu$-line is $\sim U$.
In the strict limit of $t_{\perp} = 0$, the $\nu = 1/2$ MI phase vanishes and the system is always a superfluid 
since there is a low energy path for the bosons to move along the chains.

Next, we consider the case of tilted double wells, where $\lambda \ne 0$.
The MI states present in the double-well ladders discussed here are quite different from those encountered in strictly 1D 
superlattices~\cite{rf:buonsante2}.
When $\lambda \gg {\rm max}(t_{\perp}, U)$, the MI with filling $\nu$ in the double-well ladder 
reduces to the MI with $2\nu$ in a single 1D lattice. In this regime, a transition to a SF 
phase occurs at a crictical $t_{\parallel}$
in contrast to the case of the 1D superlattice,
where all the occupied wells are completely isolated from each other and the system remains always in the MI phase.

We discuss two special cases $\lambda = U$ with $t_{\perp} \ll U$ and $\lambda \gg {\rm max}(t_{\perp}, U)$.
For $\lambda = U$ and $t_{\perp}\ll U$, where the states $|1,1\rangle$ and $|0,2\rangle$ are nearly degenerate, the width of the $\nu=1$ MI phase on the $\mu$-line is reduced to $\sim 2\sqrt{2}t_{\perp}$.
However, for $\lambda\gg {\rm max}(t_{\perp},U)$, the width of the $\nu=1$ MI phase is $\sim U$.
This happens because the $\nu = 1/2, 1$ MI phases in a double-well ladder reduce to the unit- and double-filling 
MI phases of a single 1D lattice,
as all bosons prefer to be in the lowest energy side of the largely tilted double well.

These special cases reflect the more general trend 
that as $\lambda$ increases, width of the $\nu = 1$ MI phase on the $\mu$-line changes non-monotonically.
Such non-monotonic behavior for the $\nu = 1$ MI phase is also found in $(\mu,t_{\parallel})$-plane for varying values of $\lambda$
and will be discussed next by taking into account finite $t_{\parallel}$ and studying the insulator to superfluid transition.

{\it Insulator to Superfluid Transition-I:} To include the effects of $t_{\parallel}$ and study the MI-to-SF transition, 
we use first a perturbative mean-field approach (PMFA)~\cite{rf:oosten}. 
Although PMFA fails to describe 1D systems quantitatively~\cite{rf:kuehner2}, 
it provides qualitative understanding and analytical insight.
Quantitative results can be obtained
using the TEBD method~\cite{rf:vidal1,rf:vidal2}; these results are compared
later with the picture that emerges from PMFA.

We consider the effects of finite $t_{\parallel}$ and introduce the SF order parameter 
$\psi_{\eta}=\langle a_{i,\eta}\rangle = \langle a^{\dagger}_{i,\eta}\rangle$ into 
the Hamiltonian Eq. \!(\ref{eq:hamil}), which reduces to
\begin{equation}
H \simeq \sum_i H_{i}^{\rm mf} = \sum_i [H_{i}^{\rm dw}+2t_{\parallel}\sum_{\eta}\psi_{\eta}^2 + V_{i}],
\end{equation}
where $V_i = -2t_{\parallel}\sum_{\eta}\psi_{\eta}(a_{i}^{\dagger} + a_{i})$ describes 
the transfer of atoms between $i$-th sites and the condensate $\psi_{\eta}$ and
is treated perturbatively.
  
Using perturbation theory, we obtain the correction $\Delta E_n = E_n - E^{(0)} (n,0)$
to the unperturbed ground state energy $E^{(0)}(n,0)$ in terms of $\psi_{\eta}$. 
Performing a linear transformation  $(\Psi_1,\Psi_2)^{\bf t}=X(\psi_L,\psi_R)^{\bf t}$
to diagonalize the quadratic part of $\Delta E_n$ leads to 
\begin{equation}
\label{eq:gEner1}
 \Delta E_n  =\!\! \sum_{\zeta\in\{1,2\}}\!\!\!A_{\zeta}(n,\bar{t}_{\perp},\bar{t}_{\parallel},\bar{\mu})\Psi_{\zeta}^2
  +O(\Psi_{1}^4,\Psi_{1}^3\Psi_{2},...),
\end{equation}
where $X$ is a 2 by 2 Hermitian matrix and the bars on parameters mean the normalization by $U$, e.g. $\bar{\mu}\equiv \mu/U$.
Expressions for the coefficients of $\Psi_{\zeta}^2$ and fourth order terms are quite long, thus we will not 
give them here. However, $A_2$ is always positive, while $A_1$ changes sign, and the fourth order
coefficients are positive, leading to second-order phase transitions between the MI $(\Psi_1 = \Psi_ 2 = 0)$ 
and SF $(\Psi_1 \ne 0, \Psi_2 = 0)$ states.

For symmetric double wells ($\lambda = 0$), we obtain analytical expressions for the MI-SF phase boundary 
($A_1 = 0$) in two limits.
When $t_{\perp}\ll U$, the phase boundaries are
  \begin{eqnarray}
  \bar{t}_{\parallel}^{\rm pb}&\simeq& \left\{\begin{array}{cc}
  \frac{\bar{t}_{\perp}^2-\bar{\mu}^2}{4\bar{t}_{\perp}},\,\,\,\,\, n=1 \,\,\, (\nu = 1/2),
  \vspace{2mm}
  \\
  \frac{(\bar{\mu}-\bar{t}_{\perp})(-\bar{\mu}+1-2\bar{t}_{\perp})}
  {2(\bar{\mu}+1)}, \,\,\, n=2 \,\,\, (\nu = 1),
  \end{array}\right.
  \end{eqnarray}
In this case, the critical value of $t_{\parallel}$  is 
  \begin{eqnarray}
  t_{\parallel}^{\rm c}\simeq\left\{\begin{array}{cc}
  \frac{1}{4}t_{\perp}, \,\,\,\,\,\,\,\,n=1 \,\,\, (\nu = 1/2),
  \vspace{2mm}
  \\
  \frac{3-2\sqrt{2}}{2}U-\frac{1}{2}t_{\perp},\,\,\,n=2 \,\,\, (\nu = 1),
  \end{array}\right.\label{eq:critp2}
  \end{eqnarray}
with $\mu_c \simeq 0$ for $\nu = 1/2$ and $\mu_c \simeq (\sqrt{2}-1) U$ for $\nu = 1$.
In the case of $t_{\perp}\gg U$, the double-well system reduces effectively to a single 1D lattice, 
and the phase boundaries as well as the value $t_{\parallel}^{\rm c}$ can be obtained from 
the standard results~\cite{rf:oosten} by replacing $U \to U/2$. 

%

\begin{figure}[tb]
\includegraphics[width=7.5cm, height=7cm]{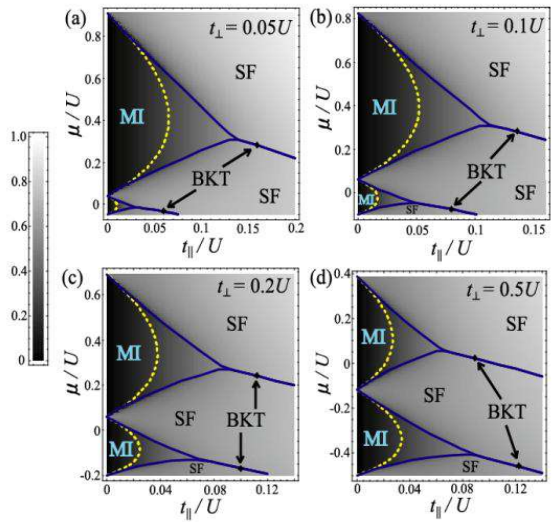}
\caption{\label{fig:PDsym}
Phase diagrams for symmetric double wells.
Dashed lines represent the phase boundary calculated within PMFA.
Solid lines, density plots, and dots are calculated by the infinite-TEBD.
$\tilde{n}$ is integer inside the solid lines.
The density plots represent $\sigma$.
The dots represent the critical point of the BKT transition.
}
\end{figure}

\begin{figure}[tb]
\includegraphics[width=6.6cm, height=4cm]{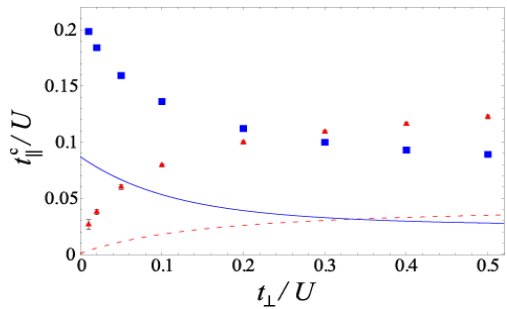}
\caption{\label{fig:tparaC}
Critical intra-chain hopping $t_{\parallel}^{\rm c}$ for symmetric double wells as a function of $t_{\perp}$.
Dashed and solid lines represent $t_{\parallel}^{\rm c}$'s for the half- and unit-filling MI phases calculated within PMFA.
Triangles and squares represent $t_{\parallel}^{\rm c}$'s for the half- and unit-filling MI phases calculated by the infinite-TEBD.
}
\end{figure}

\begin{figure}[tb]
\includegraphics[width=7.5cm, height=7cm]{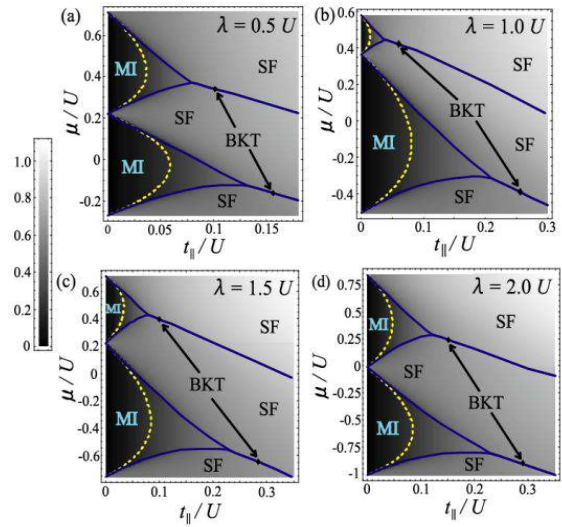}
\caption{\label{fig:PDtilt}
Phase diagrams for tilted double wells ($t_{\perp}=0.1 U$).
}
\end{figure}

\begin{figure}[tb]
\includegraphics[width=6.6cm, height=4cm]{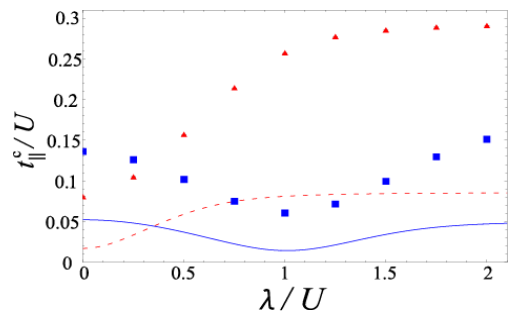}
\caption{\label{fig:tparaCtilt}
$t_{\parallel}^{\rm c}$'s for $t_{\perp}=0.1\, U$ as a function of $\lambda$ (Error bars are smaller than the size of symbols).
}
\end{figure}

In Fig. \!\ref{fig:PDsym},
the MI-SF phase boundaries for $\lambda = 0$ calculated within PMFA are shown as dashed lines
for different values of $t_{\perp}$. Notice that the figures are not in the same scale.
In Fig. \!\ref{fig:tparaC}, the critical intra-chain hoppings $t_{\parallel}^{\rm c}$'s for the $\nu = 1/2, 1$ MI phases are shown as functions of $t_{\perp}$ as dashed and solid lines.
The $\nu = 1/2$ $(\nu = 1)$ MI lobe grows (shrinks) since $t_{\parallel}^{\rm c}$ increases (decreases) with increasing $t_{\perp}$
so that the double-well system reduces to a single 1D system in the limit of $t_{\perp}\gg U$.

Next, we discuss the case of tilted double wells $(\lambda \ne 0)$.
In Fig. \!\ref{fig:PDtilt}, we show the MI-SF phase boundaries for different values of $\lambda$ 
at fixed $t_{\perp} = 0.1\, U$ indicated by dashed lines. In Fig. \!\ref{fig:tparaCtilt}, we show
the critical intra-chain hoppings $t_{\parallel}^{\rm c}$'s versus $\lambda$ for the $\nu = 1/2, 1$ MI phases 
indicated by dashed and solid lines, respectively.
The $\nu = 1/2$ MI lobe or $t_{\parallel}^{\rm c}$ grows monotonically as $\lambda$ increases.
In contrast, the $\nu = 1$ MI lobe or $t_{\parallel}^{\rm c}$ changes non-monotonically as a function of $\lambda$.

This non-monotonic behavior for $\nu = 1$ can be understood as follows:
at $\lambda=0$, $t_{\parallel}^{\rm c}$ is given approximately by  Eq. \!(\ref{eq:critp2}) 
since $t_{\perp}\ll U$ when $t_{\perp} = 0.1\, U$.
As $\lambda$ increases, $t_{\parallel}^{\rm c}$ initially decreases.
At $\lambda=U$, $t_{\parallel}$ reaches a minimum when $t_{\parallel}^{\rm c}\simeq \sqrt{2}t_{\perp}/6$,
since the states $|1,1\rangle$ and $|0,2\rangle$ states are nearly degenerate,
i.e., the local state of the MI phase at $\lambda=U$ is $\vert \Phi_{2,0}\rangle \simeq (|1,1\rangle+|0,2\rangle)/\sqrt{2}$. 
The development of this local coherence then pushes the system into the SF phase.
Further increase of $\lambda$ moves the system away from this degeneracy which favors SF,
and forces $t_{\parallel}^{\rm c}$ to increase causing a reentrance into a MI phase.
In particular, when $\lambda\gg U$ all atoms move to a single chain and are in $|0,2\rangle$,
thus the critical value becomes $t_{\parallel}^{\rm c}\simeq (5-2\sqrt{6})U/2$ as expected for a single chain~\cite{rf:oosten}.

The non-monotonic behavior of $t_{\parallel}^{\rm c}$ shows a reentrant quantum phase transition from MI to SF to MI, 
induced by the tilt $\lambda$ when $t_{\parallel}^{\rm c}$ is kept between $(t_{\parallel}^{\rm c})_{\rm min}$ and
$(t_{\parallel}^{\rm c})_{\rm max}$. 
Taking into account the high degree of control achieved in double-well optical lattices~\cite{rf:jenni}, we expect
this reentrance to be observed experimentally.
However, since we do not expect PMFA to give quantitatively correct results for the double-well (ladder) optical lattice, 
we next discuss numerical results using the TEBD method.

{\it Insulator to Superfluid Transition-II:} To determine quantitatively the phase diagrams for 
double-well (ladder) optical lattices, we use the infinite-size version of TEBD~\cite{rf:vidal2}, 
which provides an excellent ground state for 1D quantum lattice systems via imaginary time evolution.
To apply the TEBD method to our problem, we map the double-well (ladder) Bose-Hubbard model into
a single chain with {\it next-nearest}-neighbor hopping, whose ground state can be calculated via the 
{\it swapping} technique~\cite{rf:shi}.
We note that the infinite-TEBD algorithm has been recently applied to single chains with {\it only nearest}-neighbor hopping, 
where the quantum Berezinskii-Kosterlitz-Thouless (BKT) critical point~\cite{rf:zakrzewski} was obtained for the unit-filling case.
While the maximum number of bosons per site is $N_{\rm max}=\infty$, convergence is already 
achieved in our numerical calculations, when $N_{\rm max} = 4$ for $\nu = 1/2$ and $N_{\rm max} = 5$ for $\nu = 1$.

The phase diagrams in the $(\mu,t_{\parallel})$-plane are shown in Figs. \!\ref{fig:PDsym} and \ref{fig:PDtilt}, 
where the solid lines indicate the MI-SF phase boundaries,
which have roughly a triangular shape.
The sides of the MI lobe, the phase transition occurs from a $\nu = 1/2, 1$ MI to a SF with $\nu \ne 1/2, 1$. 
However, the two sides of the ``triangle'' merge for each MI phase (see dots in Figs. 2 and 4)
producing a phase transition from a $\nu = 1/2, 1$ MI to a SF with $\nu = 1/2, 1$, which is of the BKT type~\cite{rf:donohue,rf:kuehner2}.

To locate the phase boundaries we calculate directly the mean number of atoms per double well
$\tilde{n}\equiv \sum_{\eta}\langle\hat{n}_{i,\eta}\rangle$, but we also calculate the
fluctuation $\sigma\equiv \sqrt{\langle \hat{n}_i^2\rangle-\langle \hat{n}_i\rangle^2}$, 
which is small deep in the MI regions, and relatively large in the SF regions.
Since we are interested in local observables, such as $\tilde{n}$ and $\sigma$, 
convergence is already achieved for $\chi = 15$, where $\chi$ is the size of the basis set retained 
in the TEBD procedure~\cite{rf:vidal1}.

We locate the BKT transition on the lines of integer $\tilde{n}$ $(\nu = 1/2, 1)$ 
by calculating the correlation function $\langle\hat{\alpha}_{r}^{\dagger}\hat{\alpha}_0\rangle$, 
where $\hat{\alpha}_{i}^{\dagger}$ creates an atom in the lowest single particle state of a double-well.
The SF phase of the double-well ladder can be regarded as a two-band Tomonaga-Luttinger liquid (TLL),
and the correlation function exhibits power-law decay as $\langle\hat{\alpha}_{r}^{\dagger}\hat{\alpha}_0\rangle\propto r^{-K/2}$.
The exponents $K_{\rm c}$ at the phase transitions can be calculated from the TLL theory.
For instance, when $\max (t_{\perp},\lambda)\gg U$, our system is effectively a single 1D chain
and has the critical value $K_{\rm c}=1/2$ for the BKT transition~\cite{rf:kuehner2}. In addition, 
when $\lambda=0$ and $\tilde{n}=2$ $(\nu = 1)$, the critical value is also $K_c = 1/2$~\cite{rf:donohue}. 
Consequently we use the criterion $K_{\rm c}=1/2$ to identify the critical point for the BKT transition at integer 
values $\tilde{n}$ $(\nu = 1/2, 1)$.

We calculate $K$ as a function of $t_{\parallel}$ 
by fitting $a\cdot r^{-K/2}$ to the correlation function calculated from the TEBD method with $\chi=60$. 
We use the intervals $10\le r\le 15$, $15\le r\le 20$, $20\le r\le 25$, and $25\le r\le 30$ for the fit and take 
the average of them to produce error bars. The critical intra-chain hopping $t_{\parallel}^{\rm c}$ along the 
lines of integer $\tilde{n}$ is determined when $K = K_c$. The dots in Figs. \!\ref{fig:PDsym} and \ref{fig:PDtilt}
correspond to the BKT transition points.
In Figs. \!\ref{fig:tparaC} and \ref{fig:tparaCtilt}, $t_{\parallel}^{\rm c}$'s for $\tilde{n}=1,2$ $(\nu =1/2, 1)$ 
are shown as triangles and squares, respectively.
The phase boundaries asymptotically approach those of PMFA as $t_{\parallel}$ tends to zero.
On the other hand, differences between PMFA and TEBD are significant when $t_{\parallel}$ is relatively large. 
In particular, $t_{\parallel}^{\rm c}$ obtained using TEBD is more than twice as large as that obtained within PMFA.
However, the qualitative behavior of the phase diagram as a function of $t_{\perp}$ and $\lambda$ 
obtained within PMFA is consistent with that of the infinite-TEBD.

{\it Conclusions:} In summary, we have studied the superfluid-to-Mott insulator transition of bosons in double-well (ladder) optical lattices.
Applying the time-evolving block decimation (TEBD) method to the two-leg Bose-Hubbard model, we have calculated the zero-temperature phase diagram.
We have found that the phase diagram changes significantly depending on the inter-chain hopping and tilt of the double wells.
In particular, we have shown that the tilt can be used to induce 
reentrant transitions between Mott insulator and superfluid phases.
Through a comparison of the results of TEBD and the perturbative mean-field approach (PMFA),
we have shown that PMFA fails to describe the phase diagram quantitatively, but captures 
its qualitative trends.

We acknowledge support from a Grant-in-Aid from JSPS (I. D.) and from 
NSF-DMR-0304380 (C. SdM.).

\end{document}